\documentclass[]{spie}  

\usepackage{amsmath,amsfonts,amssymb}
\usepackage{graphicx}
\usepackage{siunitx}
\usepackage[colorlinks=true, allcolors=blue]{hyperref}
\usepackage[acronym]{glossaries}
\usepackage{subcaption}

\title{Assembly, integration, and verification of MITESI: an optical testbed for simulating the ELT in the laboratory}

\author[a]{Daniel J Ahrer}
\author[a]{M. Concepci\'{o}n C\'{a}rdenas V\'{a}zquez}
\author[a]{Thomas Bertram}
\author[a]{Peter Bizenberger}
\author[a]{Armin Huber}
\author[a]{Rapha\"{e}l Pourcelot}
\author[a]{Silvia Scheithauer}
\author[a]{Horst Steuer}

\affil[a]{Max Planck Institute for Astronomy, K\"{o}nigstuhl 17, 69117, Heidelberg, Germany}

\authorinfo{Send correspondence to D.A.\\D.A.: E-mail: danahrer@mpia.de}

\newacronym{ao}{AO}{Adaptive Optics}
\newacronym{cad}{CAD}{Computer Aided Design}
\newacronym{cfo}{CFO}{Common Fore Optics}
\newacronym{cmo}{CMO}{Command Matrix Optimiser}
\newacronym{dm}{DM}{Deformable Mirror}
\newacronym{elt}{ELT}{Extremely Large Telescope}
\newacronym{fov}{FoV}{Field of View}
\newacronym{fp}{FP}{Focal Plane}
\newacronym{fwhm}{FWHM}{Full Width Half Maximum}
\newacronym{hci}{HCI}{High Contrast Imaging}
\newacronym{metis}{METIS}{Mid-Infrared ELT Imager and Spectrograph}
\newacronym{mitesi}{MITESI}{MIni elt TElescope SImulator}
\newacronym{mpia}{MPIA}{Max Planck Institute for Astronomy}
\newacronym{ngs}{NGS}{Natural Guide Star}
\newacronym{og}{OG}{Optical Group}
\newacronym{pp}{PP}{Pupil Plane}
\newacronym{ppc}{PPC}{Pupil Position Control}
\newacronym{psf}{PSF}{Point Spread Function}
\newacronym{rms}{RMS}{Root Mean Square}
\newacronym{scao}{SCAO}{Single Conjugate Adaptive Optics}
\newacronym{wfe}{WFE}{WaveFront Error}
\newacronym{wfs}{WFS}{WaveFront Sensor}
 
\begin{document} 
\maketitle

\begin{abstract}
MITESI is an optical testbed which simulates key characteristics of the ELT, ultimately producing an artificial natural guide star. It was designed primarily to enable testing of the METIS instrument's SCAO system in closed loop. In this contribution, we discuss the assembly, integration and verification process of the testbed, taking the project from design to assembled hardware.
\end{abstract}


\section{Introduction} \label{sec:intro}

\gls*{mitesi} is an \gls*{ao} testbed designed to simulate a \gls*{ngs} observed with the \gls*{elt}. Its primary use case is to test the \gls*{scao} system \cite{2024SPIE13097E..27B,2024ExA....58...20F} of the \gls*{metis} instrument \cite{2021Msngr.182...22B}. This will occur in two stages, the first will be subsystem testing of the \gls*{scao} system alone at the \gls*{mpia}, where the exit \gls*{fp} of \gls*{mitesi} will interface to the entrance \gls*{fp} of the \gls*{scao} module, the wavefront sensor of the \gls*{scao} system. The second stage of testing at system level will see \gls*{mitesi} installed in front of \gls*{metis} in Leiden.

The top level requirement for \gls*{mitesi} is to enable testing of wavefront sensing and closed loop operations of the \gls*{scao} system during subsystem and system testing, mimicking operations at the \gls*{elt}. This drives a number of the required functionalities of \gls*{mitesi} including the need to simulate a diffraction limited point source (with the goal of also being able to simulate an extended object), and to be able to interface this point source in the same manner that the \gls*{scao} module will receive a natural guide star at the \gls*{elt}. As well as replicating the \gls*{fp}, a pupil with similar characteristics to the \gls*{elt} pupil is also necessary. Testing closed loop operations also requires the ability to inject spatial and time varying wavefront errors to simulate the atmospheric seeing, among other things, as well as the ability to apply the wavefront corrections derived from the \gls*{scao} module. For this purpose \gls*{mitesi} contains a \gls*{dm}. \gls*{mitesi} will also be used to test the \gls*{scao} system's ability to sense misregistration \cite{2024SPIE13097E..4LS} and lateral motion of the pupil image on the \gls*{wfs} \cite{coppejans:hal-04440483,2024ExA....58...20F}. To test these, \gls*{mitesi} can laterally shift the pupil image projected onto the \gls*{dm}, and the pupil image downstream of the \gls*{dm}. Finally, the shortest wavelength \gls*{mitesi} is required to emit at is $\lambda$~=~\SI{1.5}{\micro\metre} and the longest $\lambda$~=~\SI{4.1}{\micro\metre}. The shorter wavelengths are required for testing of the \gls*{scao} system which operates in the astronomical H and K bands, whereas the longer wavelengths are covered to enable testing of \gls*{metis} \gls*{hci} modes, which will occur during system level testing\cite{2024SPIE13096E..52A}. More details about the requirements of \gls*{mitesi} can be found in \cite{Conchi-2026-MITESI}.

\begin{figure}
    \centering
    \includegraphics[width=\linewidth]{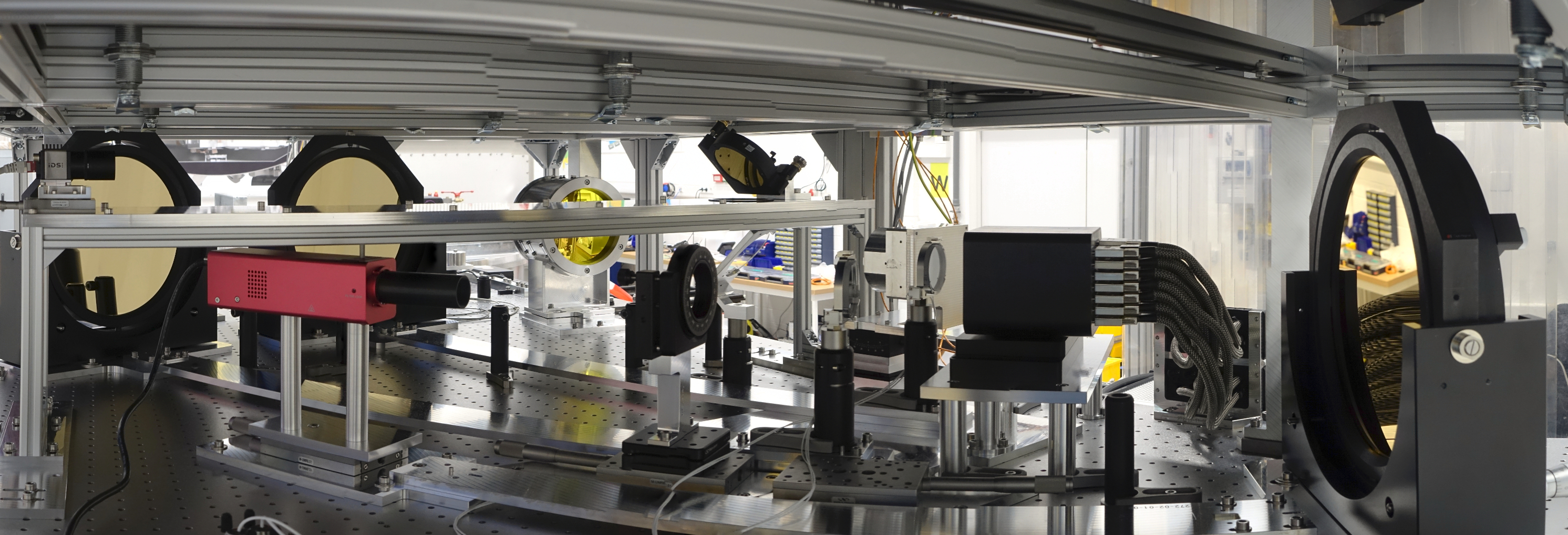}
    \vspace{1mm}
    \caption{The \gls*{mitesi} test bench installed in the cleanroom where it will be used for subsystem testing of the \gls*{scao} system. Note that normally black cover panels are attached to the aluminium profile frame. The components are labelled in figure~\ref{fig:MITESI-layout}.}
    \label{fig:MITESI-lab}
\end{figure}

Since our last proceeding \cite{2024SPIE13097E..53M}, \gls*{mitesi} has been assembled and delivered to the cleanroom where it will be used for subsystem testing of the \gls*{scao} module, a recent view of the bench is shown in figure~\ref{fig:MITESI-lab}. In this proceeding we discuss the assembly and integration process before going on to present some of the initial verification testing of the test bench to demonstrate its performance.

\section{Design overview}

\begin{figure}
    \centering
    \includegraphics[width=1\linewidth]{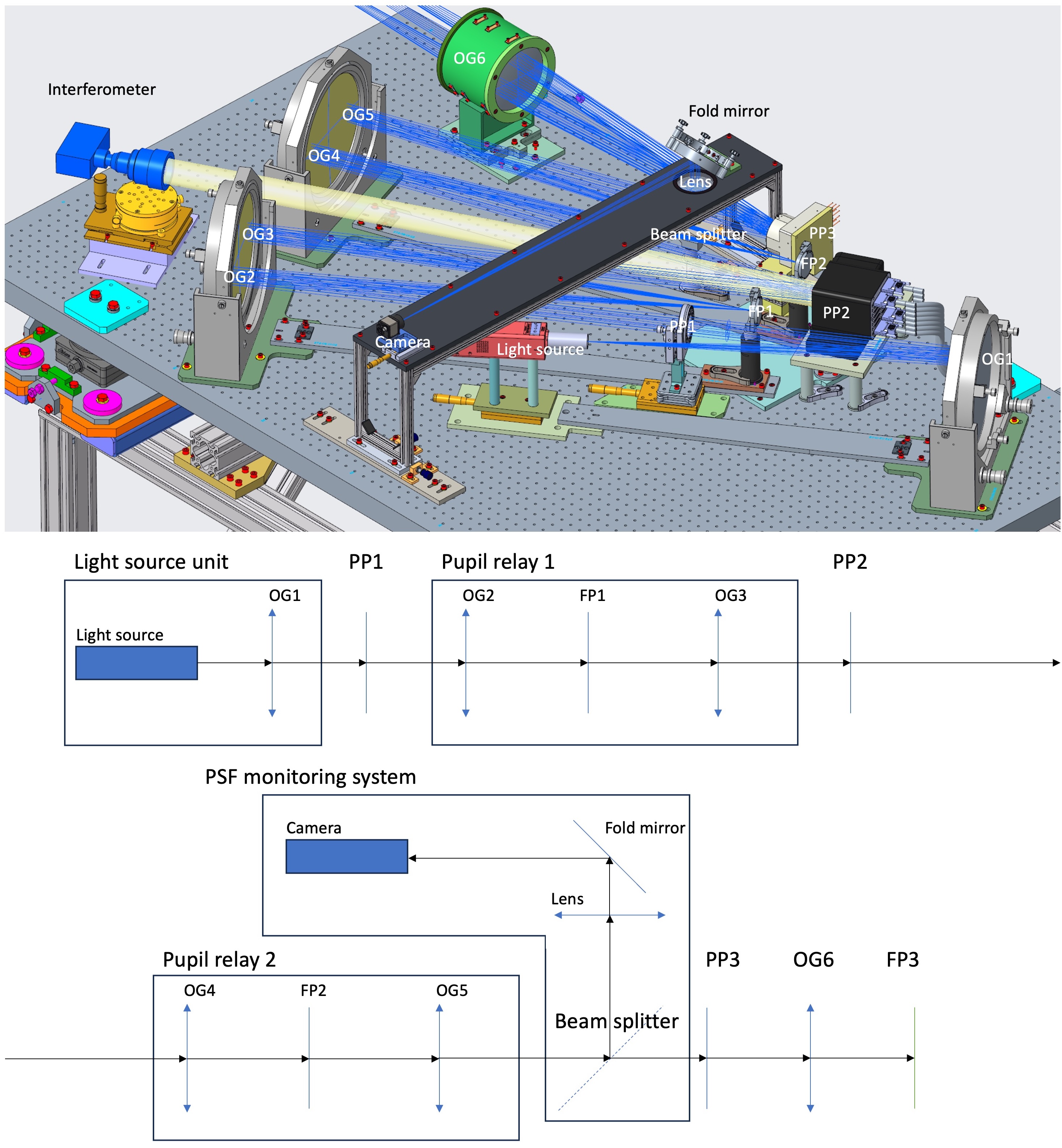}
    \vspace{1mm}    
    \caption{Top: a 3D CAD view of the \gls*{mitesi} bench in its configuration for subsystem testing at MPIA. The cover, among other items, have been hidden for clarity. Bottom: a sketch of the main optical path of \gls*{mitesi} from the light source unit to the lens unit \gls*{og}6 which forms the interface focal plane (not shown).}
    \label{fig:MITESI-layout}
\end{figure}

To achieve the required functionality, \gls*{mitesi} includes optical elements that replicate the properties of the \gls*{elt} relevant to wavefront sensing with the \gls*{scao} pyramid \gls*{wfs}. An overview of \gls*{mitesi} optical layout is given in figure~\ref{fig:MITESI-layout}. 

\gls*{mitesi} comprises of a \textbf{light source unit} which features a Thorlabs SLS203F/M light source with two Thorlabs LA5370 lenses in a (4f) configuration which reimage the emitter onto the pinhole. The standard pinhole size used in \gls*{mitesi} is \SI{25}{\micro\metre}, but this can be exchanged, for example to simulate an extended object. The light emitted by the pinhole is then reflected by \gls*{og}1, an on-axis parabola, which collimates the light and propagates it to \gls*{pp}1.

\begin{figure}
    \centering
    \includegraphics[width=0.5\linewidth]{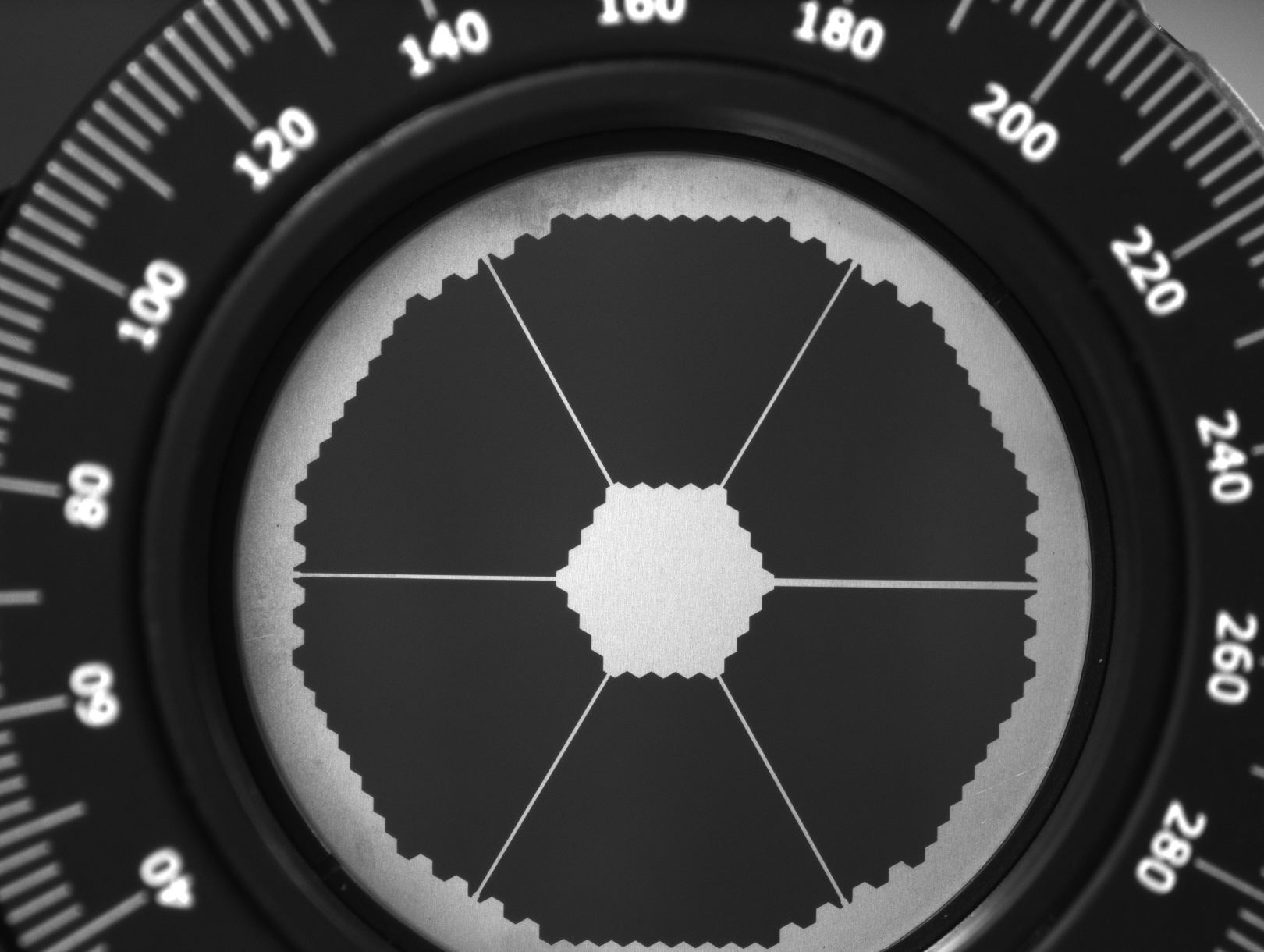}
    \vspace{1mm}
    \caption{The \gls*{mitesi} \gls*{elt} pupil mask, mounted in a manual rotation stage.}
    \label{fig:ELT-mask}
\end{figure}

At \textbf{\gls*{pp}1}, masks can be placed to simulate various pupil geometries. The default mask used is our \gls*{elt} pupil mask shown in figure~\ref{fig:ELT-mask}, which is mounted in a manual rotation stage. Two other masks are currently available, one is a simple circular aperture and the other a Bahtinov mask to aid with quick focusing. All masks are mounted on magnetic baseplates which interface to the \gls*{mitesi} bench, allowing for quick switching without the need for realignment.

\begin{figure}
    \centering
    \includegraphics[width=0.5\linewidth]{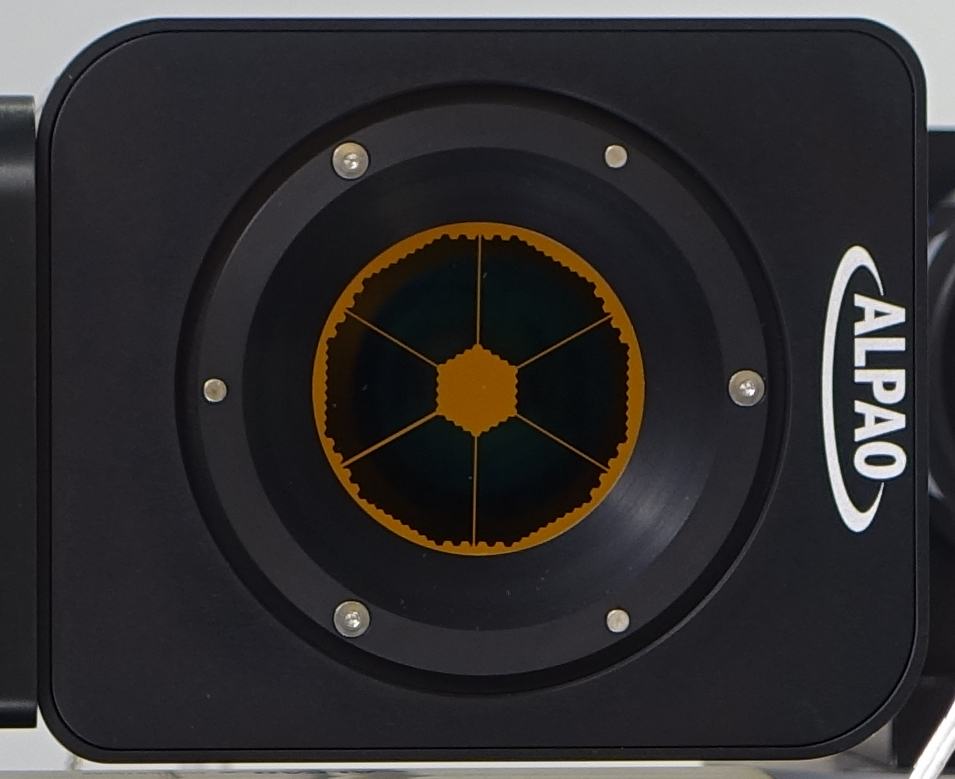}
    \vspace{1mm}
    \caption{The \gls*{elt} mask located at \gls*{pp}1, imaged onto \gls*{pp}2 through the pupil relay 1 optical group.}
    \label{fig:PP1-projected-pp2}
\end{figure}

The next optical group, \textbf{pupil relay 1}, consists of an on-axis parabolic mirror and a flat mirror placed at the focus. This group serves two primary functions, the first is the focal plane mirror, which is mounted in a motorised tip-tilt stage, allows for lateral motion of the image of \gls*{pp}1 at the subsequent pupil plane. The second function is to reimage \gls*{pp}1 to \gls*{pp}2, figure~\ref{fig:PP1-projected-pp2} shows the image of the \gls*{elt} mask at \gls*{pp}1, on the \gls*{dm} at \gls*{pp}2, taken during alignment.  

Located at \textbf{\gls*{pp}2} is an ALPAO \gls*{dm}820, featuring 820 actuators. This \gls*{dm} serves to both inject and apply corrections for atmospheric turbulence. I.e. in open loop, atmospheric turbulence will be applied to the \gls*{dm} but in closed loop operation the residual of the atmospheric turbulence minus the correction from the \gls*{scao} system will be applied to the \gls*{dm}.

The angle of incidence on the \gls*{dm} is designed such that the full aperture of the \gls*{dm} can be viewed on-axis from behind the two large on-axis parabolas. Here we place a compact interferometer so we can measure the shape of the \gls*{dm} while it is embedded in \gls*{mitesi}.

\textbf{Pupil relay 2} has the same design and functionality as pupil relay 1, in that it reimages the pupil plane at the \gls*{dm} to the next pupil plane, while also providing access to a tip-tilt actuated flat mirror in the focal plane which laterally moves the image of the previous pupil plane. 

At this point in the optical path, a small amount of the light is picked off from the main path by the \textbf{beam splitter} which is an uncoated elliptical calcium fluoride window. This light is directed vertically, through a long focal length lens attached to the long bench seen above the main optical table in figure \ref{fig:MITESI-layout}. The light is then folded by a flat mirror and comes to a focus on a camera located at the other side of the bench. This creates a well sampled \gls*{psf} which can be recorded while operating \gls*{mitesi}. 

Up to this point in the optical path, the light has only propagated on-axis but we wish to place the simulated \gls*{ngs} anywhere within the 27 arcsecond on-sky \gls*{fov} the \gls*{scao} module is designed to accept. Therefore, at \textbf{\gls*{pp}3} is a flat mirror in a tip-tilt stage. We intend to use this stage to also test motion of the point source in the \gls*{fov} during operations, for example in the case of non-sidereal tracking at the \gls*{elt} or even circular trajectories, to simulate the motion of an off-axis point source seen during observations when \gls*{metis} is used in a pupil tracking mode.

Finally, the light passes through \textbf{\gls*{og}6}, an f-theta design triplet lens which brings the light to the interface focal plane (\gls*{fp}3). It is designed so that during subsystem testing, \gls*{fp}3 replicates the interface between the \gls*{scao} module and the \gls*{metis} \gls*{cfo}, and during system testing between \gls*{metis} and the \gls*{elt}. For example by matching the beam f-ratio. \gls*{og}6 also reimages \gls*{pp}3 to replicate the \gls*{elt} exit pupil received by the \gls*{scao} module and \gls*{metis} at subsystem and system level testing respectively.

\section{Assembly and Integration} \label{sec:assembly-integration}

\gls*{mitesi} is integrated on a dedicated 1.2 x \SI{2.1}{\metre} optical table with a regular grid hole pattern. A number of components are aligned on the \gls*{mitesi} table by referencing them to an alignment plate with a series of dowel pin holes. In this section, we detail how the individual components of \gls*{mitesi} were aligned on this optical table.

\subsection{Integration of large parabolas and focal plane mirrors} \label{sec:align-og23-og34}

The two large parabolas, which together form OG2/3 and OG4/5 serve a couple optical functions, the first is that they reimage the pupil planes to the next, with OG2/3 reimaging PP1 to PP2 and OG4/5 re-imaging PP2 at PP3. They also provide access to intermediate focal planes at the foci of the parabolas. By tip-tilting a flat mirror placed at these focal planes, we can laterally shift image of the previous pupil plane, i.e. tip-tilting FP1 moves the image of PP1 projected at PP2.

Alignment of each of the large parabolas and their focal plane mirrors was done with a reference ball and an interferometer with a converging beam. The ball was placed at the approximate focus of the parabola mechanically, with its position referenced to a dowel pin. The interferometer was placed behind it, pointing towards the parabola. The spacing between the parabola and ball was then adjusted so that a null was seen from both the ball and the parabola (with the ball removed and a flat mirror placed after the parabola to return the beam in double pass) with the interferometer at the same location. With this technique, the focus of the parabola was then aligned to the centre of the reference ball to within tens of microns. The ball and its mount was then removed and replaced by the focal plane mirror assembly, which picked up the position of the reference ball by using the dowel pin the ball mount was attached to.

The parabola and focal plane unit was aligned to the rest of the optical table via a dowel pin connecting it to the alignment plate.

\subsection{Integration of DM}

The positioning of the \gls*{dm} was achieved mechanically, again via a series of dowel pins which referenced the \gls*{dm} to the alignment plate. 

The pointing of the \gls*{dm} was aligned by placing an interferometer with a collimated beam pointing towards \gls*{og}2. This beam then passed through the optical system, reflecting off the \gls*{dm}, up to \gls*{fp}2, where a reference ball was placed, returning the light to the interferometer. A flat was then applied to the \gls*{dm} and the \gls*{dm} tip-tilted in its gimbal mount to achieve an approximate null on the interferometer. 

\subsection{Conjugation of pupil planes 1 and 2}

\gls*{pp}1 is the location of our pupil plane masks. Unlike previous units, it was not referenced to the alignment plate via a dowel pin. Instead, the unit was placed on the table to within the precision of the screw holes. The unit is mounted to a Newport M-BK-3A magnetically repeatable mount so different masks can be easily swapped in and out. 

The magnetically repeatable mount sits on-top of a linear stage which enables fine adjustment of the spacing between the \gls*{pp}1 mask and \gls*{og}2 to conjugate the \gls*{pp}1 mask to the surface of the \gls*{dm}. 

Coarse conjugation was done by using a camera with a C-mount lens pointing towards \gls*{og}2, focusing the camera on the image of the \gls*{dm} mirror surface. \gls*{pp}1 was then placed in front of the camera on the magnetic mount and adjusted to also bring it into focus on the camera. 

Fine conjugation was then achieved by using a point source microscope with a shallow depth of field. With the point source microscope, defocus as small as \SI{50}{\micro\metre} is discernable at \gls*{pp}1. The point source microscope, which was also on a linear stage, was then brought to a focus on the \gls*{pp}1 mask, roughly conjugating it to the surface of the \gls*{dm}. The \gls*{pp}1 mask was then removed and the point source microscope adjusted in focus until conjugated to \gls*{pp}2, the \gls*{pp}1 mask was then placed back in front of the point source microscope and it was now adjusted to bring it to the focus of the point source microscope, meaning the \gls*{pp}1 mask and \gls*{dm} were now both in focus on the point source microscope.

This two stage approach was necessary as the depth of focus of the point source microscope was too shallow to easily blindly search for the \gls*{dm} mirror surface. Lateral alignment of the \gls*{pp}1 mask image on the \gls*{dm} was done by tip-tilting \gls*{fp}1 until the image of \gls*{pp}1 was centred on \gls*{pp}2 as seen by imaging \gls*{pp}2 with a camera.

\subsection{Integration of PP3 and OG6}

\gls*{pp}3 is a tip-tilt mirror placed in a pupil plane, up until now the beam path of \gls*{mitesi} has been on-axis, however for testing the \gls*{scao} system, we need to be able to place the artificial natural guide star anywhere within the \gls*{fov} it accepts. 

To enable this functionality a mirror is placed in a motorised tip-tilt mount in a pupil plane. As the range of this motorised tip-tilt stage is not sufficient to cover the entire \gls*{scao} module \gls*{fov}, it is supplemented with a second manual stage consisting of a rotation and goniometric stage. 

\gls*{og}6 is a triple lens system which serves two primary purposes. Firstly, it converts the collimated beam reflecting off of \gls*{pp}3 into a focal plane at the exit focal plane \gls*{fp}3. Secondly, it reimage the the pupil plane at \gls*{pp}3 to create the exit pupil plane \gls*{pp}4.

\begin{figure}
    \centering
    \includegraphics[width=0.5\linewidth]{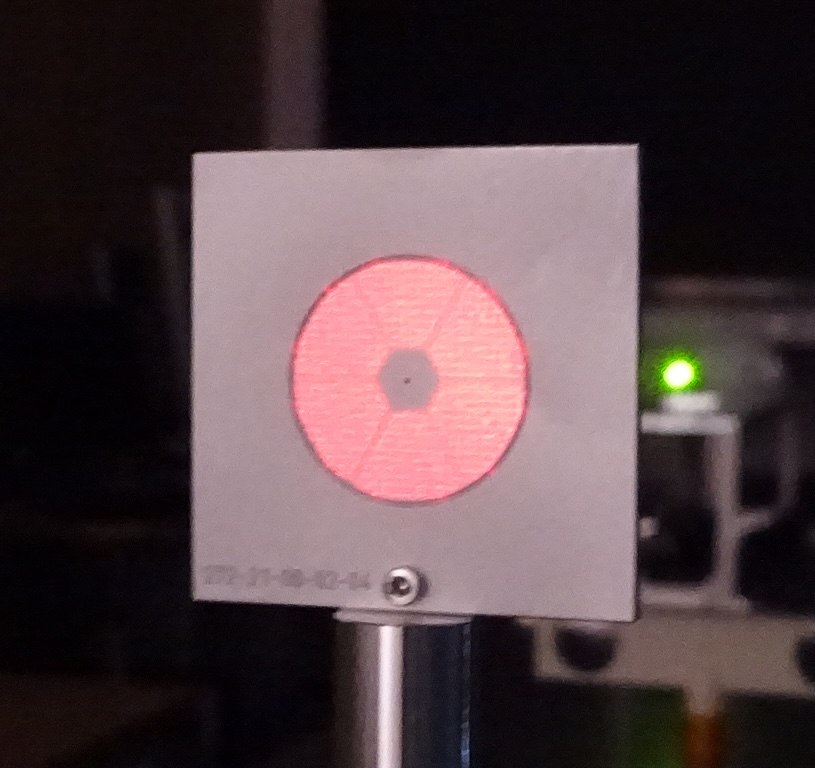}
    \vspace{1mm}
    \caption{The footprint of the \gls*{elt} mask centred on an alignment target placed at \gls*{fp}3. The dot at the centre of the ring was used to verify the pointing of \gls*{og}6 at a later alignment stage.}
    \label{fig:PP3-tip-tilt-align}
\end{figure}

To integrate \gls*{pp}3, its stack of stages was mounted to the central interface plate by two dowel pins which set both the position on the optical table and the initial pointing. Final alignment of the pointing was done by placing a target with a ring, the centre of which was marked with a small dot, at the expected position of \gls*{fp}3. The manual stages of \gls*{pp}3 were then adjusted until laser light from an interferometer pointing along the main beam path of \gls*{mitesi}, and reflecting off \gls*{pp}3, was centred on the alignment target. An example of this is shown in figure \ref{fig:PP3-tip-tilt-align}. The \gls*{og}6 lens unit was then placed on the optical table, a lens cap with a circular indent at its centre was placed over the side of the lens facing \gls*{pp}3, the x position of the lens was adjusted until the lens cap indent was centred on the beam reflected from \gls*{pp}3. The height of the lens unit was fixed and set by the mechanical design of the lens unit mount but was also verified by observing that the indent in the lens cap was centred on the beam reflected from \gls*{pp}3.

The spacing between \gls*{pp}3 and \gls*{og}6 was set with a ruler running from the edge of the interface plate. A knife edge square translated the surface of the \gls*{og}6 lens barrel to the optical table and the position \gls*{og}6 was adjusted until the spacing between the lens barrel and interface plate was as per the \gls*{mitesi} \gls*{cad} model.

The tilt of the \gls*{og}6 lens unit was verified by observing that the focus of the lens unit was aligned with a dot centred on the ring on the \gls*{fp}3 target that \gls*{pp}3 was previously aligned to.

\subsection{Integration of light source}

The light source unit consists of a light source and a large on-axis parabola. The source is a modified Thorlabs SLS203F/M with two LA5370 lenses in a 4f configuration, re-imaging the silicon carbide globar emitter onto a pinhole. The light source unit is designed such that this pinhole is then at the focus of the \gls*{og}1 parabola so that the parabola collimates the light and steers it towards \gls*{pp}1. 

The focus of the parabola was aligned in the same way as the \gls*{og}2/3 and \gls*{og}4/5 parabolas, discussed in section \ref{sec:align-og23-og34}, i.e. an interferometer with a converging lens was placed behind the parabola focus and a reference ball on a dowel pinned mount was placed at the mechanically expected focus of the parabola. The spacing between the ball and parabola was then adjusted until the centre of the ball and the focus of the parabola coincided. 

\begin{figure}
    \centering
    \includegraphics[width=0.8\linewidth]{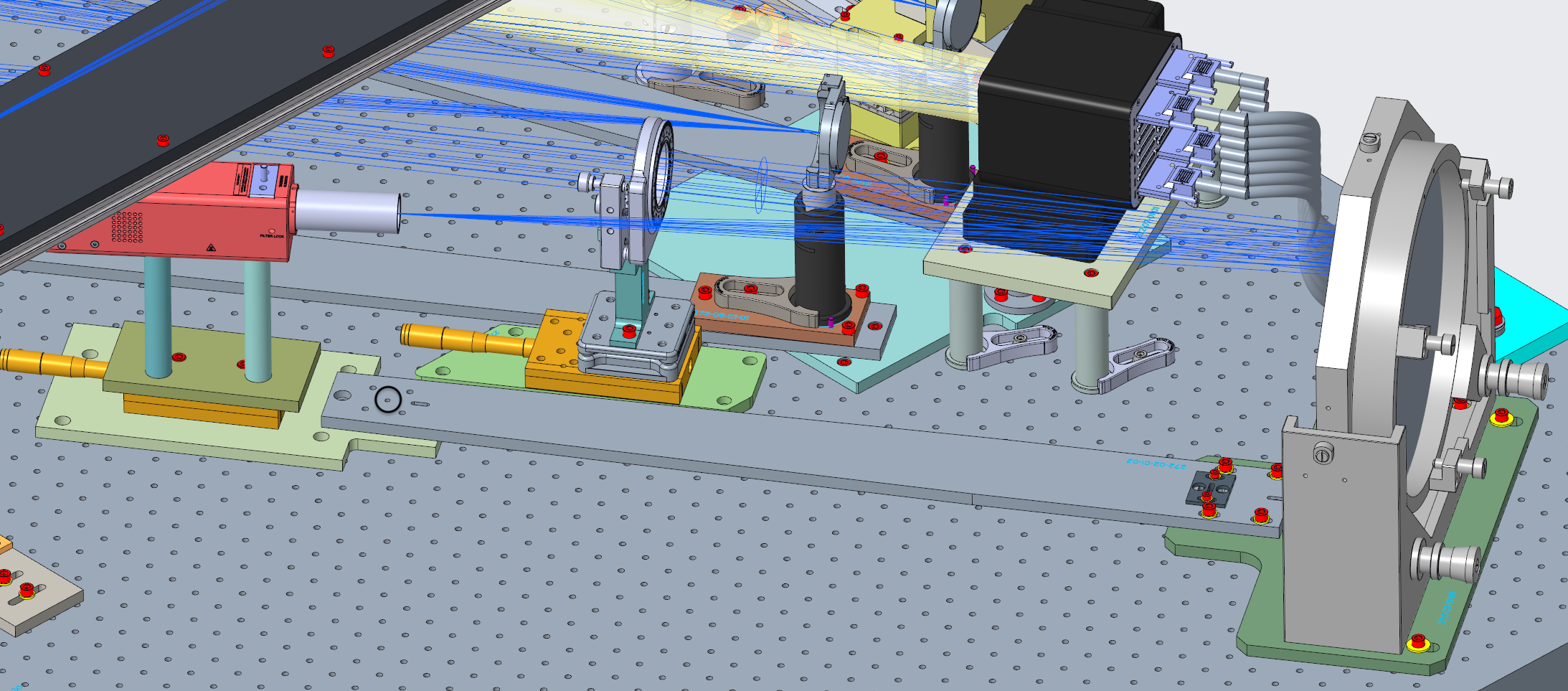}
    \vspace{1mm}
    \caption{The light source unit in the \gls*{mitesi} \gls*{cad} file. The dowel-pin hole at the focus of the \gls*{og}1 mirror is marked by the black circle.}
    \label{fig:OG1-dowel-pin}
\end{figure}

Here, it was also necessary to ensure the collimated beam after reflection from \gls*{og}1 was on-axis (i.e. propagates perpendicular to the surface of the \gls*{pp}1 mask). To enable this, the mounting for the \gls*{og}1 parabola can swing about the dowel pin at its focus (see figure~\ref{fig:OG1-dowel-pin}). To align the parabola in rotation, the \gls*{fp}1 mirror assembly was removed from the bench and a reference ball mounted at the focus of the \gls*{og}2/3 parabola was installed instead. The light from an interferometer with the converging lens was then collimated by \gls*{og}1 and propagated towards \gls*{og}2, reflecting off the ball at the focus of \gls*{og}2, and returning to the interferometer. \gls*{og}1 was then rotated about the dowel pin to minimise the tilt of the returned the interference fringes.

With \gls*{og}1 aligned, the interferometer was removed and replaced with the light source. The pointing and XY position of which is set by the two screw holes in the plate the light source is attached to. The z position, distance to the parabola, was set by referencing the emitter nose piece to the dowel pin at the focus of \gls*{og}1.

\subsection{Integration of the PSF monitoring system}

There are two main units of the \gls*{psf} monitoring system. The first is the beam splitter, which is mounted in such a way that it is held at a 45 degree angle and connected to a \SI{25}{mm} pillar post, inserted into a post holder. This assembly is then secured to the optical table with a clamping fork. As the beam splitter is a flat optic in a collimated beam, the positioning in XY and Z is not critical optically. However the XY positioning (in the plane of the optical table) does matter as there is only a few millimetres of spacing between the beam splitter and its mount on one side with the rays that trace the pupil re-imaging and on the other with the rays that trace the one of the off-axis pointings after \gls*{pp}3. The XY position of the beam splitter was set by centring the beam splitter on the footprint of the light from the light source after reflection from \gls*{og}5. 

The \gls*{psf} monitoring system bench was then aligned to the beam splitter. The centration of the bench to the beamsplitter was done by attaching a C-mount lens to the PSF monitoring system camera so that instead of the image plane, the pupil plane was imaged onto the camera. With this, it was possible to make a composite image of the pupil and the edge of the lens barrel and adjust the position of the bench to centre the lens over the pupil image. 

With the C-mount lens removed from the camera, the \gls*{psf} was acquired on the camera. The camera was brought to focus using a linear stage underneath it, taking images of the \gls*{psf} at different positions of the linear stage and fitting the \gls*{fwhm} to identify by visual inspection of the focus curve, the optimal focus.

Unfortunately the \gls*{psf} currently observed on the \gls*{psf} monitoring system camera is broader than expected and shows significant aberrations when defocused. This is being investigated but we believe the aberrations originate in the \gls*{psf} monitoring system as we see a good quality \gls*{psf} at the exit focal plane of \gls*{mitesi}, see section~\ref{sec:PSF-comparison}.

\section{Verification}

In this section, we present preliminary results demonstrating the performance of \gls*{mitesi}.

\subsection{WFE measurements} \label{sec:WFE-measurement}

We targeted a total \gls*{wfe} through the whole system of $\leq$\SI{110}{\nano\metre} \gls*{rms}, when applying only a flat to the \gls*{dm}, i.e. not using the \gls*{dm} for \gls*{wfe} compensation. 

To verify the alignment and test bench, we aimed to make a \gls*{wfe} measurement using an interferometer. For the most part this is not an issue when using the interferometer at the $\lambda$ = \SI{633}{\nano\metre} measurement wavelength, as most components are reflective and so pass a broad range of wavelengths. This is an issue for the triple lens group \gls*{og}6 however, which shows very little transmission at $\lambda$ = \SI{633}{\nano\metre}. It is possible to see a faint focused spot at the exit of \gls*{og}6 but it is so faint we have not tried to make a measurement through the lens group with the interferometer. Instead, we have made \gls*{wfe} measurements though the majority of the \gls*{mitesi} system. In one configuration, this was done by using a converging lens on the interferometer, placing its focus at the focus of \gls*{og}1, where the pinhole is normally located, and placing a flat mirror downstream in collimated parts of the \gls*{mitesi} beam path to return the light in double pass. We also measured the \gls*{wfe} by placing an interferometer with a collimated beam in front of \gls*{og}6, sending the light backwards and reflecting off either a flat mirror placed in collimated space in \gls*{mitesi} or a reference ball, placed at the focus of \gls*{og}1.

Initial attempts to simply measure from the focus of \gls*{og}1 to just in front of \gls*{og}6 were made difficult by a high \gls*{wfe} from the \gls*{dm}, of order \SI{200}{\nano\metre}, despite placing a flat on the mirror. This was later found to be mostly coming from the fact the \gls*{dm} drifts while holding the flat, with the \gls*{wfe} degrading the longer the flat is held. 

We attempted to remove the \gls*{wfe} contribution of the \gls*{dm} by pointing an interferometer directly at it, measuring it in isolation and subtracting this in quadrature from the measurements made through the main \gls*{mitesi} beam path measured at locations after the \gls*{dm}. This was unsuccessful as the \gls*{wfe} measurement value of the \gls*{dm} alone was slightly larger than the \gls*{wfe} measurement through the rest of \gls*{mitesi}. 

\begin{figure}
    \centering
    \includegraphics[width=0.8\linewidth]{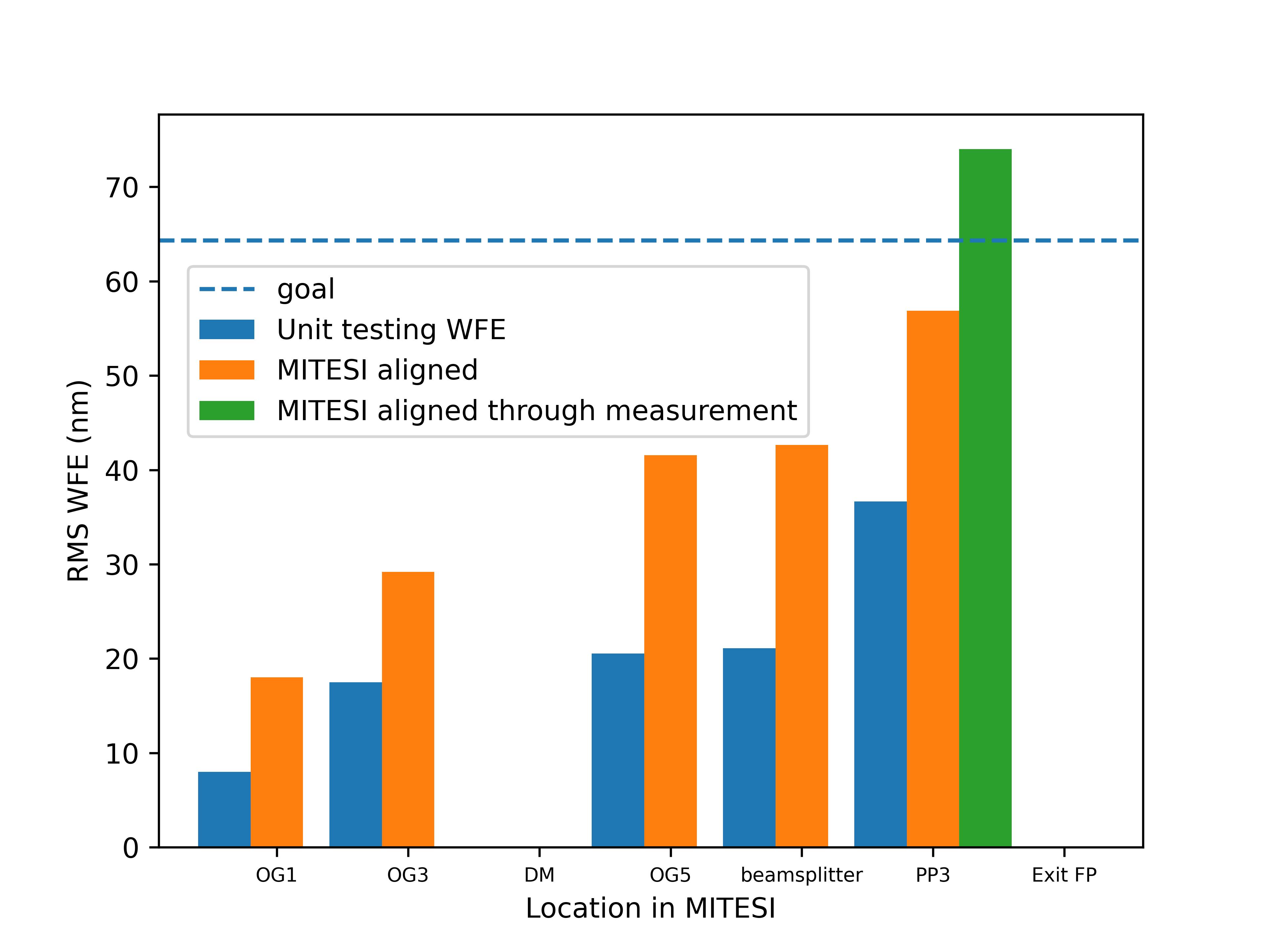}
    \vspace{1mm}
    \caption{A comparison of the cumulative \gls*{wfe} measured through \gls*{mitesi} and the running total of the \gls*{rms} \gls*{wfe} adding the measurements of the individual optics taken during unit testing. The ``\gls*{mitesi} aligned" bar represents the measurements made by making measurements in both the forwards and backwards directions as explained in the text. ``\gls*{mitesi} aligned through measurement" is the single measurement made from in front of \gls*{og}6 going backwards through the system to a reference ball placed at the focus of \gls*{og}1.}
    \label{fig:MITESI-WFE}
\end{figure}

To work around the high \gls*{wfe} from the \gls*{dm}, \gls*{wfe} measurements were made from the focus of \gls*{og}1 up to the \gls*{dm} and from in front of \gls*{og}6, backwards through \gls*{mitesi}, again to in front of the \gls*{dm}. By combining these two data sets it is possible to estimate the \gls*{wfe} from the focus of \gls*{og}1 to the just before \gls*{og}6, excluding the contribution from the \gls*{dm}. This is shown in figure~\ref{fig:MITESI-WFE} by the ``\gls*{mitesi} aligned" bars where the measurement labelled ``\gls*{og}3" for example, represents the \gls*{wfe} from the location of the light source pinhole to after reflection off of \gls*{og}3, the path of which is explained in figure~\ref{fig:MITESI-layout}.

In addition to the ``\gls*{mitesi} aligned" dataset which represents the combination of measurements made in both directions, we were also able to make a measurement backwards through \gls*{mitesi}, placing the interferometer in front of \gls*{og}6 and a reference ball at the location of the light source pinhole, measuring the \gls*{wfe} of the system, including the \gls*{dm}, in one measurement. This is represented by the green ``\gls*{mitesi} aligned through measurement" bar. This was taken at a later date when the impact of holding a flat on the \gls*{dm} was understood.

Finally, for comparison, we also include the estimated \gls*{wfe} at the various points in \gls*{mitesi} from combining in quadrature the \gls*{wfe} measured for the individual optics during their testing before integration into \gls*{mitesi}. This comparison should not be seen as one to one comparison, but an order of magnitude, as there are a number of factors that don't make the data directly comparable. For one, the data for the individual optics were taken with different interferometers and sometimes the power term was removed, sometimes it wasn't. The footprint used to represent the \gls*{mitesi} beam was sometimes slightly different and not necessarily centred on the exact part of the mirror actually used within \gls*{mitesi}. In these calculations we also assume that \gls*{og}3 contributes the same \gls*{wfe} as \gls*{og}2 as they are reflections off the same mirror, though from different sections of the mirror, the same is assumed for \gls*{og}4 and \gls*{og}5. Finally, while the \gls*{wfe} of the focal plane mirrors was measured during unit testing it is not included in this running unit testing \gls*{wfe} as the footprint of the \gls*{mitesi} beam is so small the contribution of the focal plane mirrors is assumed to be negligible. 

The dashed ``goal" line represents terms extracted from the \gls*{mitesi} \gls*{wfe} budget relevant to compare to the laboratory measurements, specifically the on-axis \gls*{rms} \gls*{wfe} before \gls*{og}6 from the optical design of \gls*{mitesi}, the allocation from manufacturing of the optics and the allocation for optomechanical assembly/integration. 

As figure~\ref{fig:MITESI-WFE} shows, both the \gls*{wfe} estimated from the unit testing and from measuring the optics without the influence of the \gls*{dm} are below the \gls*{wfe} goal. Unfortunately, the measurement made in one go from in front of \gls*{og}6 to the ball at the focus of \gls*{og}1, ``MITESI aligned through measurement" exceeds the goal by around \SI{10}{\nano\metre}. This can, at least partially, be explained by what appears to be a couple of actuators that were stuck on the \gls*{dm} at the time of the measurement, creating a significant \gls*{wfe} around them. 

While we cannot say if we have met our overall \gls*{wfe} goal of \SI{110}{\nano\metre}, as this goal includes the contribution from \gls*{og}6 which we have not been able to measure, we believe \gls*{og}6 is performing well as we are able to observe a good quality \gls*{psf}s at the exit focal plane, as we discuss in section~\ref{sec:PSF-comparison}.

\subsection{PSF comparison} \label{sec:PSF-comparison}

\begin{figure}
    \centering
    \begin{subfigure}[b]{0.49\textwidth}
        \centering
        \includegraphics[width=\linewidth]{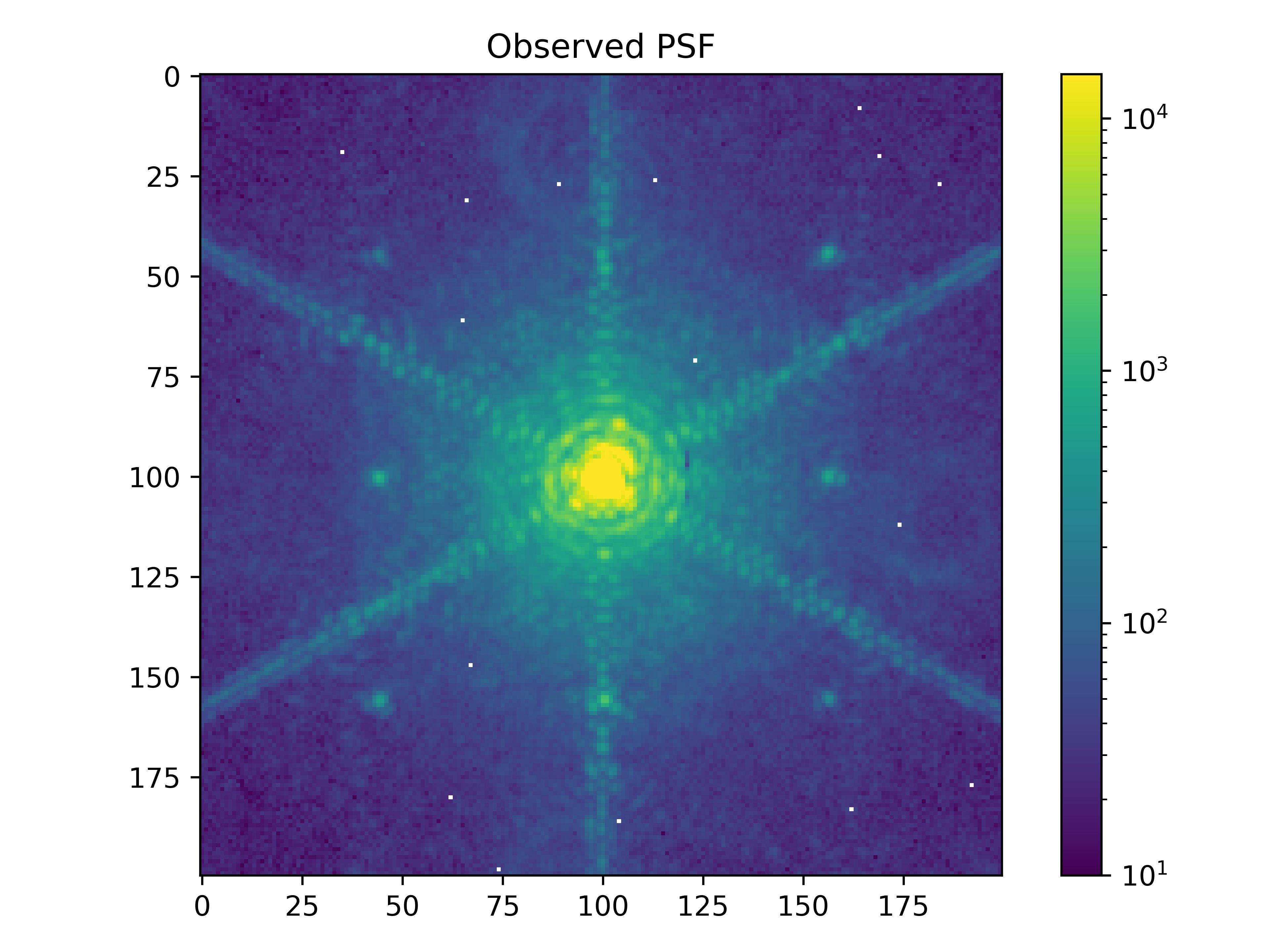}
        \label{fig:observed-psf}
    \end{subfigure}
    \hfill
    \begin{subfigure}[b]{0.49\textwidth}
        \centering
        \includegraphics[width=\linewidth]{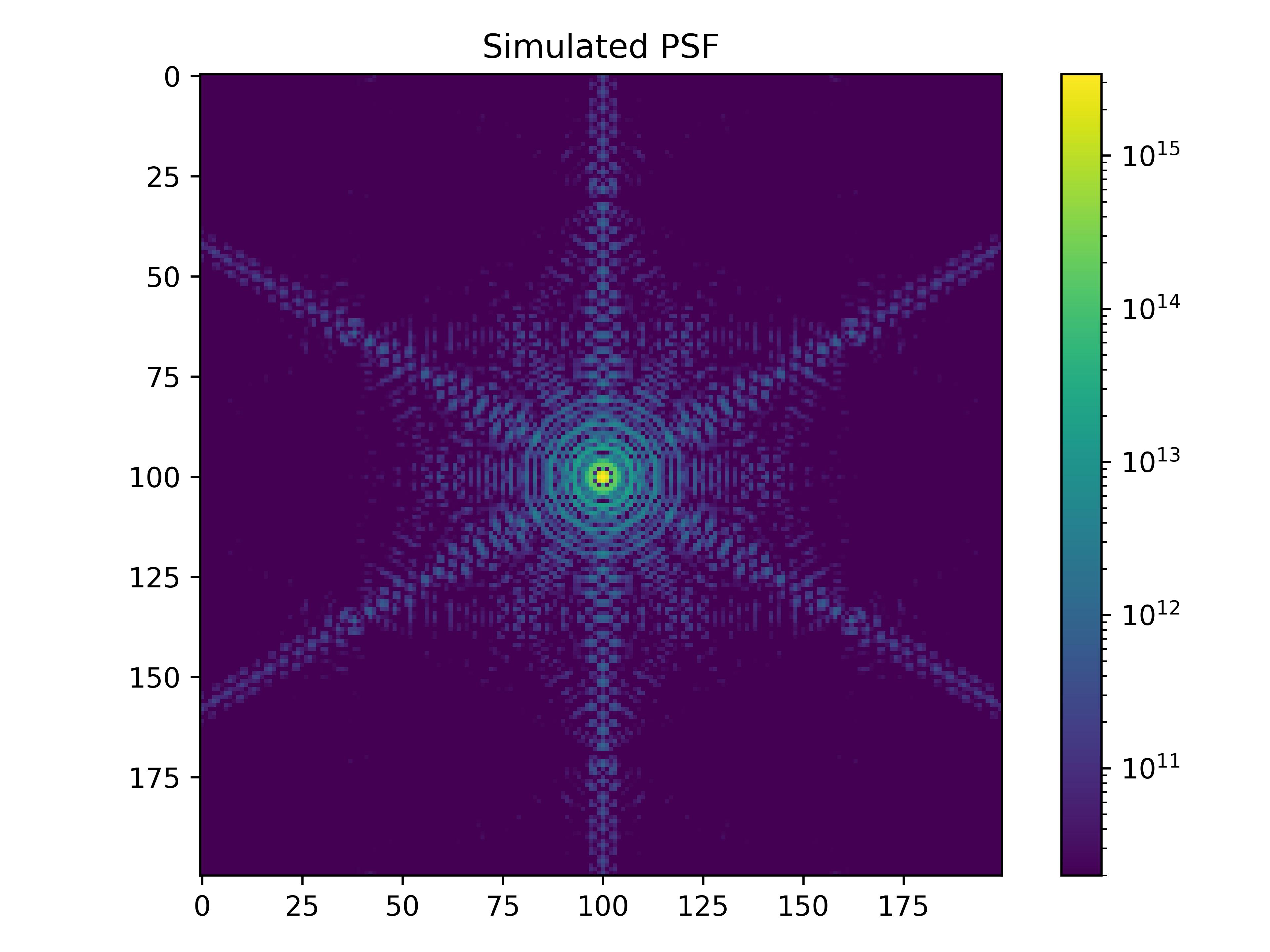}
        \label{fig:simulated-psf}
    \end{subfigure}
    \vspace{1mm}
    \caption{Left: The \gls*{psf} at the exit focal plane (\gls*{fp}3) of \gls*{mitesi} on a log scale. Shown here is the median of 5 frames that has been corrected by a dark frame. Note that in this \gls*{psf} the core is saturated. Right: A \gls*{psf} simulated at $\lambda$ = \SI{1.646}{\micro\metre} and sampled on \SI{15}{\micro\metre} pixels to match that of the observed data.}
    \label{fig:MITESI-PSF}
\end{figure}

The interface between both the \gls*{elt} and \gls*{metis} and the \gls*{scao} module to \gls*{metis} is a focal plane. Hence the interface of \gls*{mitesi} is also a focal plane, designed to replicate the properties (f-ratio, plate scale, distance to pupil) \gls*{metis} and the \gls*{scao} module will interface to. While this focal plane will not be accessible during operation of the testbed, during the verification process we have installed an extension breadboard and placed a C-RED 2 camera at the interface focal plane. An example \gls*{psf} taken with a narrowband filter ($\lambda$ = \SI{1.646}{\micro\metre}, $\Delta\lambda$ = \SI{19.51}{\nano\metre} (\gls*{fwhm})) placed in the light source can be seen in figure~\ref{fig:MITESI-PSF}. It contains the familiar characteristics of the \gls*{elt} \gls*{psf} including the diffraction spikes. One interesting property is the faint grid of dots seen in the image plane. The spacing of the dots is ~56 pixels or \SI{840}{\micro\metre} with the C-RED 2 pixel pitch. The cause of this grid in the focal plane is the \SI{1.5}{\milli\metre} pitch of the actuators of the \gls*{dm} in the pupil plane.

To quantify the quality of the \gls*{psf} at \gls*{fp}3 we compare it to a simulated \gls*{elt} \gls*{psf}. The simulated \gls*{psf} is monochromatic at $\lambda$ = \SI{1.646}{\micro\metre} and sampled on \SI{15}{\micro\metre} pixels to match the pixel pitch of our lab data, it is shown on the right of figure~\ref{fig:MITESI-PSF}. 

With this simulated \gls*{psf} we are able to calculate the Strehl ratio of the observed \gls*{psf} \cite{2004SPIE.5490..504R}. The first step is to process the \gls*{psf} observed at the exit of \gls*{mitesi} by subtracting a median dark from from the individual light frames, before combining them into a median light frame. The simulated \gls*{psf} is then rotated to match the orientation of the diffraction spikes in the observed \gls*{psf}. The simulated \gls*{psf} is then normalised to have the same counts as the observed \gls*{psf} within a 30 pixel wide square. This width was chosen to capture as much of the observed \gls*{psf} before the signal is dominated by the background while also in this case cutting out four pixels which had negative values after subtracting the dark frame. 

\begin{figure}
    \centering
    \begin{subfigure}[b]{0.49\textwidth}
        \centering
        \includegraphics[width=\linewidth]{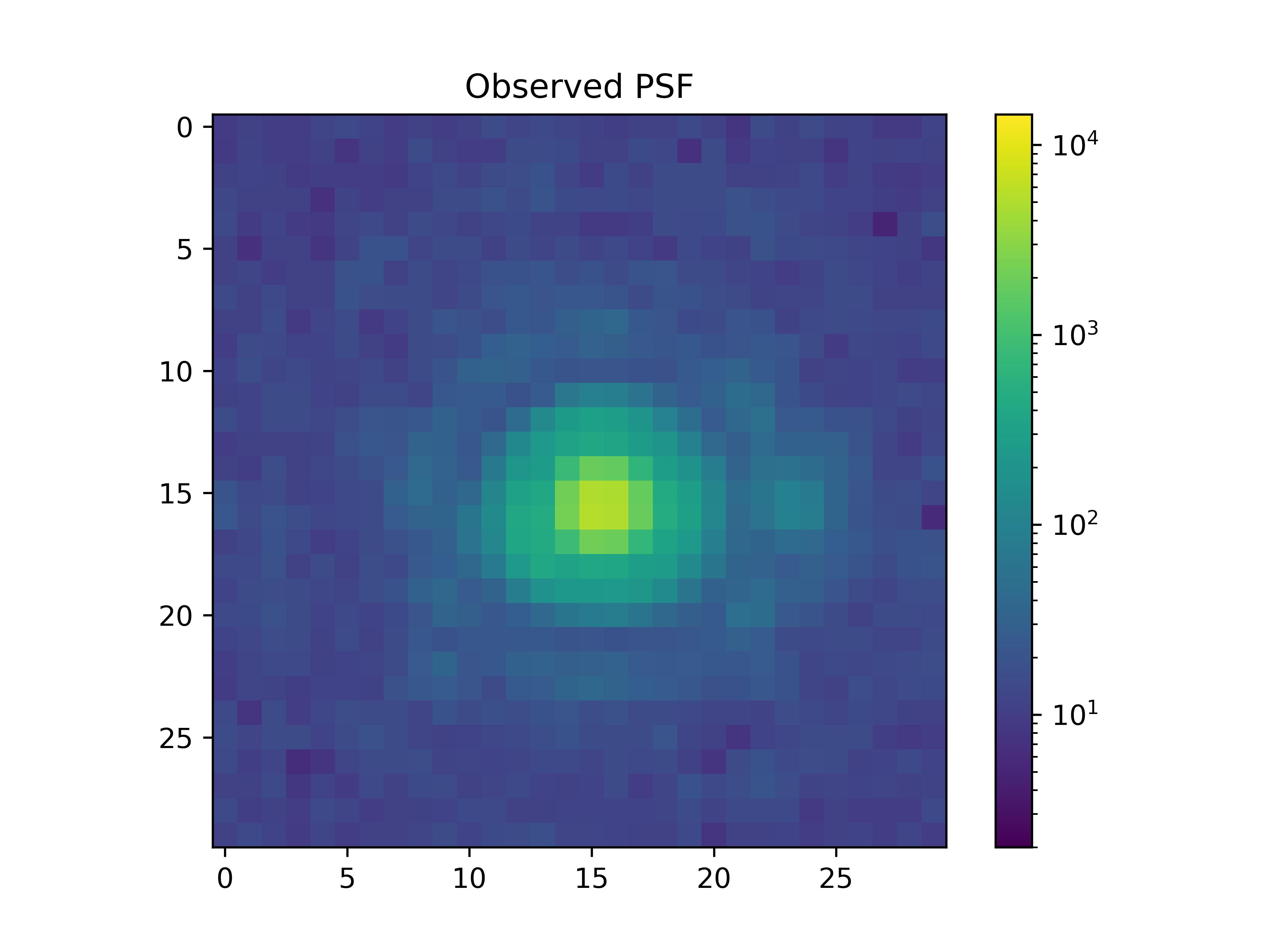}
    \end{subfigure}
    \hfill
    \begin{subfigure}[b]{0.49\textwidth}
        \centering
        \includegraphics[width=\linewidth]{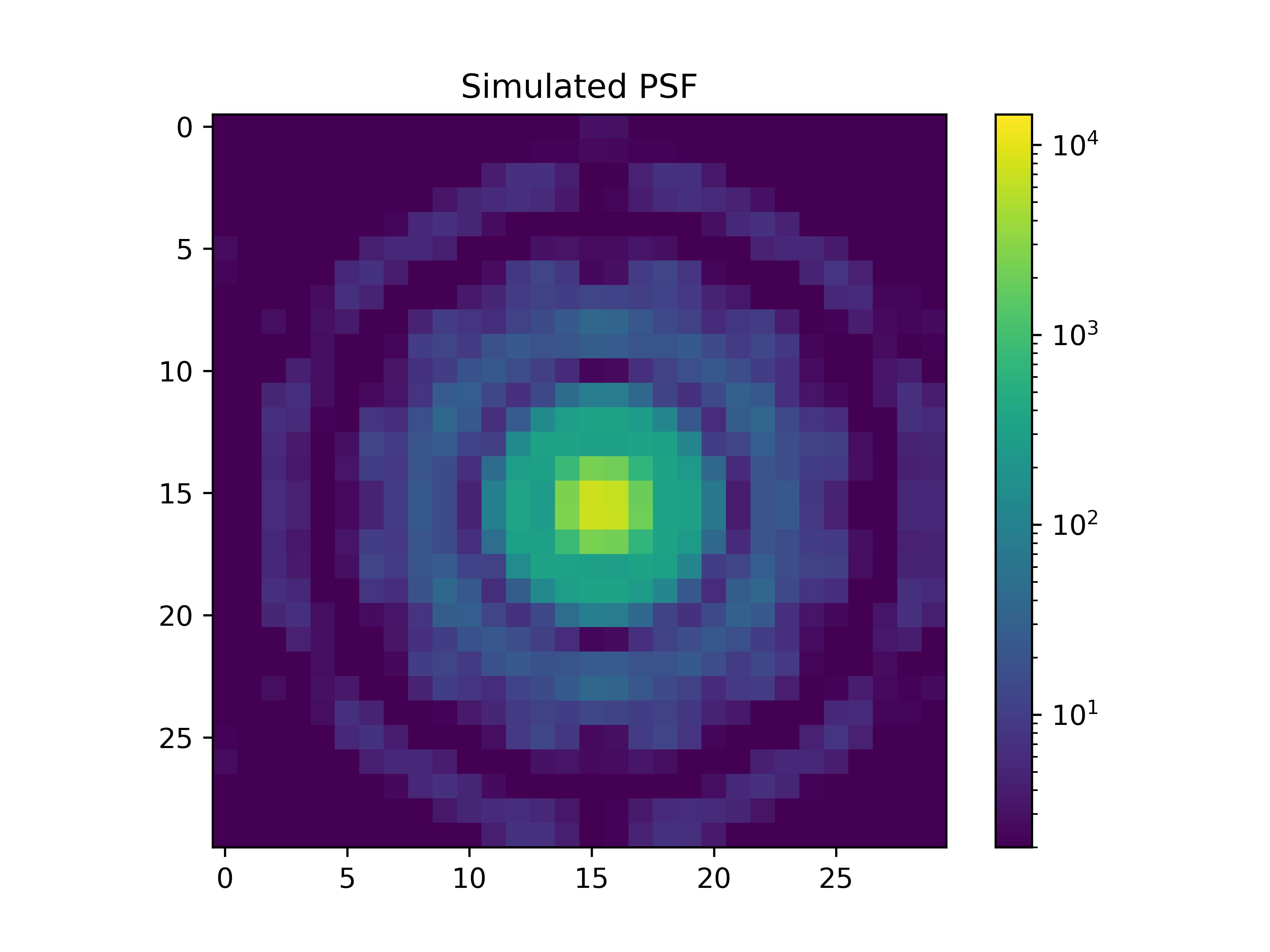}
    \end{subfigure}
    \begin{subfigure}[b]{0.49\textwidth}
        \centering
        \includegraphics[width=\linewidth]{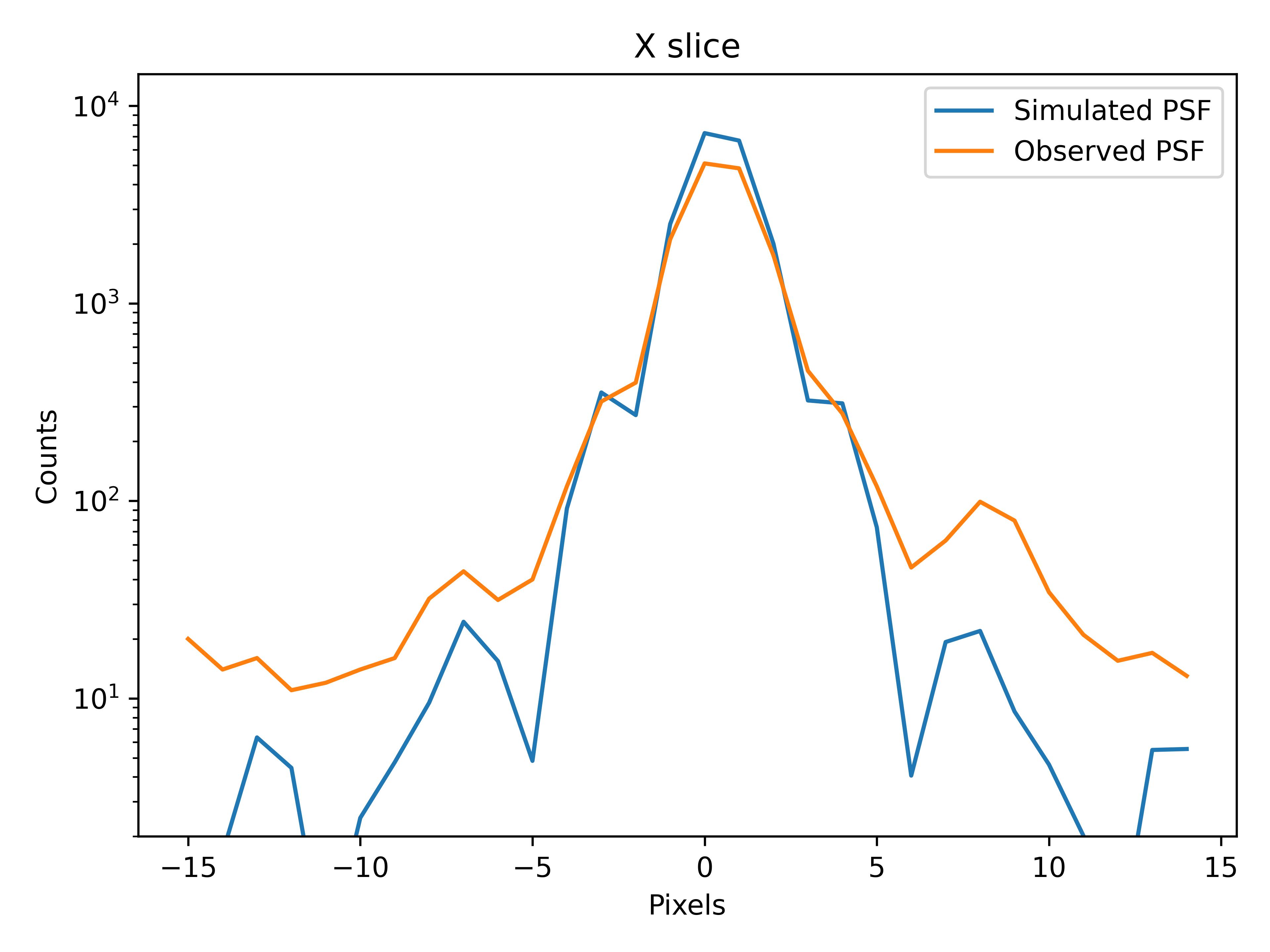}
    \end{subfigure}
    \hfill
    \begin{subfigure}[b]{0.49\textwidth}
        \centering
        \includegraphics[width=\linewidth]{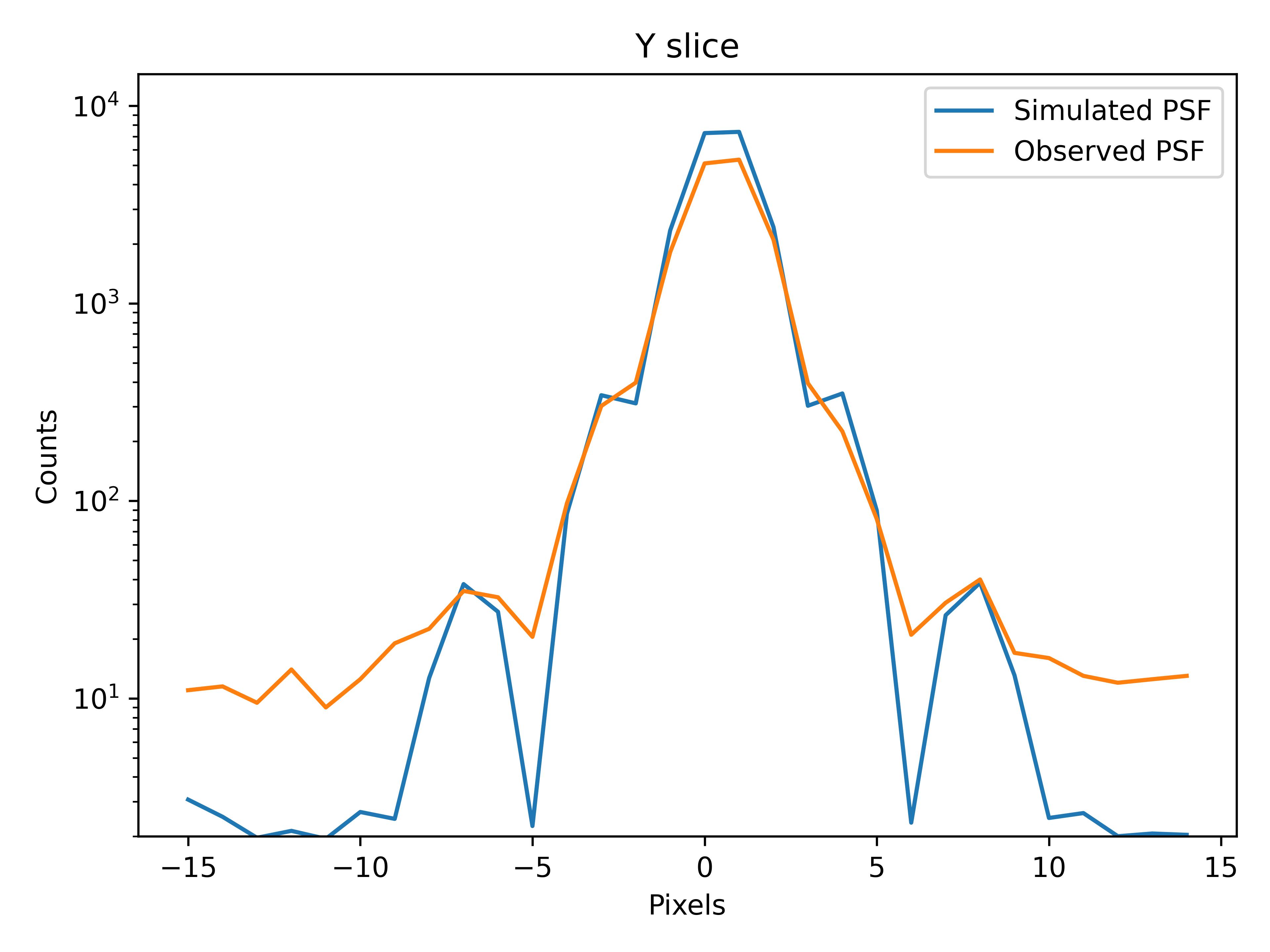}
    \end{subfigure}
    \begin{subfigure}[b]{0.49\textwidth}
        \centering
        \includegraphics[width=\linewidth]{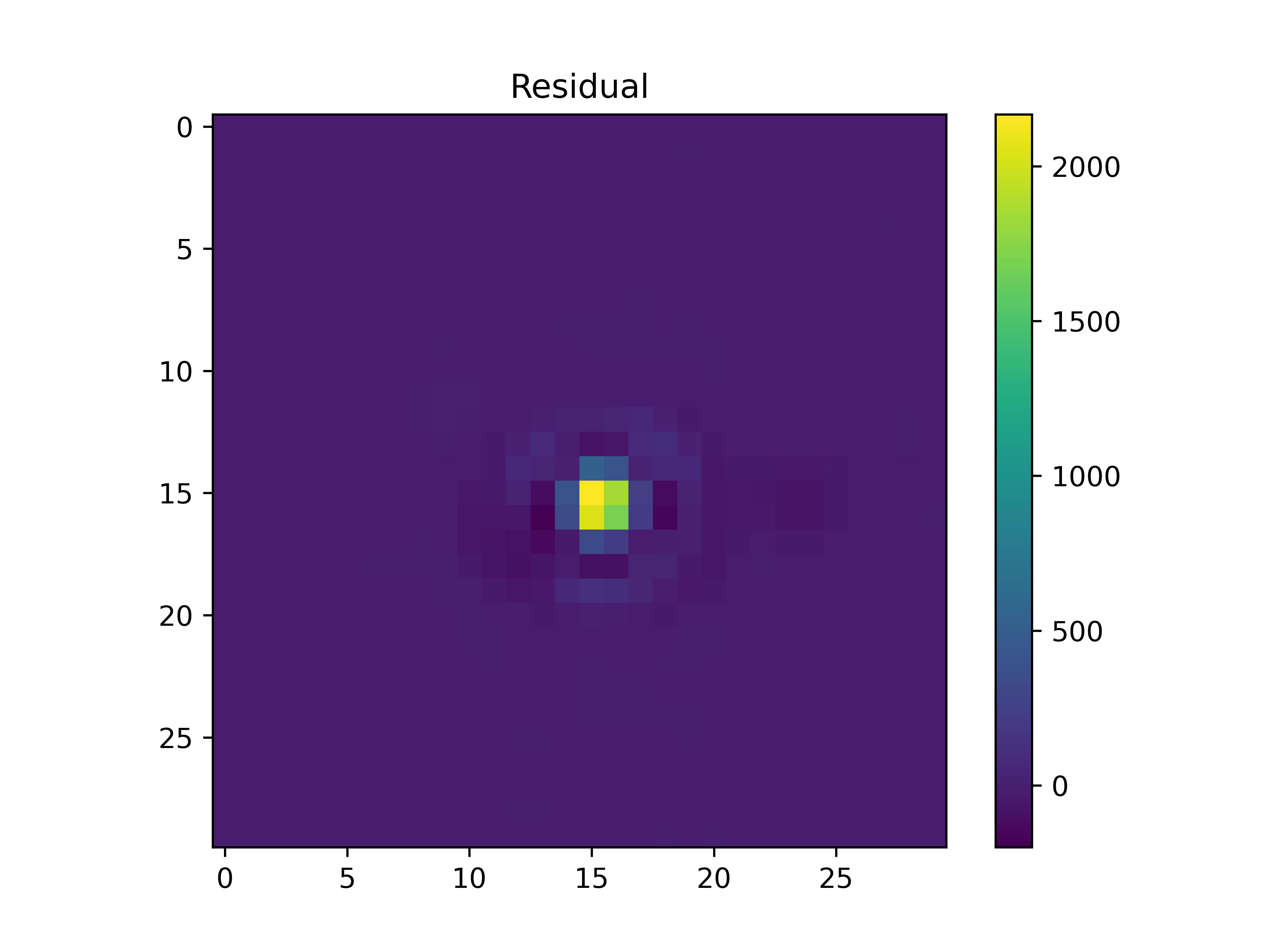}
    \end{subfigure}
    \caption{A comparison of the \gls*{psf}s used to calculate the strehl ratio. The top row shows the observed \gls*{psf} on the left, and the simulated \gls*{psf} after normalisation and sub-pixel shifting to match the observed \gls*{psf} on the right. Note the observed \gls*{psf} shown here is different to the one in figure~\ref{fig:MITESI-PSF}. The middle row shows X and Y 1D slices of the simulated and observed \gls*{psf}s. The bottom row is the residual of the simulated \gls*{psf} minus the observed \gls*{psf}, here on a linear scale to display negative values.}
    \label{fig:MITESI-Strehl}
\end{figure}

As the core of the \gls*{psf} appears to be spread over multiple pixels in our observed data, whereas it is centred on a single pixel in our simulated \gls*{psf}, the simulated \gls*{psf} is shifted at a sub pixel level to match the profile of the observed \gls*{psf}. The offset between the simulated and lab \gls*{psf} is calculated by cross correlation with the scikit-image phase cross correlation routine, before the simulated \gls*{psf} is then shifted using the scipy.ndimage.shift routine with the order of the spline interpolation changed from the default value of three, which produced negative intensities in the shifted \gls*{psf}, to one. Turning off the prefilter led to a blurring of the shifted simulated \gls*{psf}, dropping the peak intensity and hence increasing the calculated Strehl ratio. The cross correlation and shifting of the simulated \gls*{psf} was done twice, as after shifting it once it was found that the pixel with the highest intensity in the simulated \gls*{psf} was not the same as in the observed \gls*{psf}. Repeating the cross correlation and shifting process resolved this. The processed observed and simulated \gls*{psf}s are shown in the top panel of figure~\ref{fig:MITESI-Strehl}. After this, the ratio of the peak pixel intensity of the observed \gls*{psf} to the processed simulated \gls*{psf} is taken, resulting in a Strehl ratio of 0.72. This Strehl is lower than we would expect, given that the extended Mar\'echal approximation for a \gls*{rms} \gls*{wfe} of \SI{74}{\nano\metre} (excluding \gls*{og}6) presented in \ref{sec:WFE-measurement} would produce a Strehl of 0.92. The discrepancy can at least partially be explained by two factors. The first, that the \gls*{wfe} presented in \ref{sec:WFE-measurement} does not include the contribution of \gls*{og}6, whereas the \gls*{psf} measured at \gls*{fp}3 passes through the lens. The second is that even after applying dark corrections there are still around 10 counts per pixel in the background of the observed \gls*{psf}. This pushes up the total counts used to normalise the simulated \gls*{psf}, increasing the intensity of the simulated \gls*{psf}, reducing the Strehl ratio.

One caveat is that the \gls*{elt} pupil from which the \gls*{psf} is generated, both for our simulated and real \gls*{psf}, is rotationally asymmetric as one of the spiders is thicker than the other five. As we do not currently know the orientation of the \gls*{psf} recorded with \gls*{mitesi} with respect to the simulated \gls*{psf} there could be a mismatch but at the resolution of the C-RED 2 pixels it's hard to see any asymmetry in the \gls*{psf}. 

\section{Conclusions and future work}

\begin{figure}
    \centering
    \includegraphics[width=0.8\linewidth]{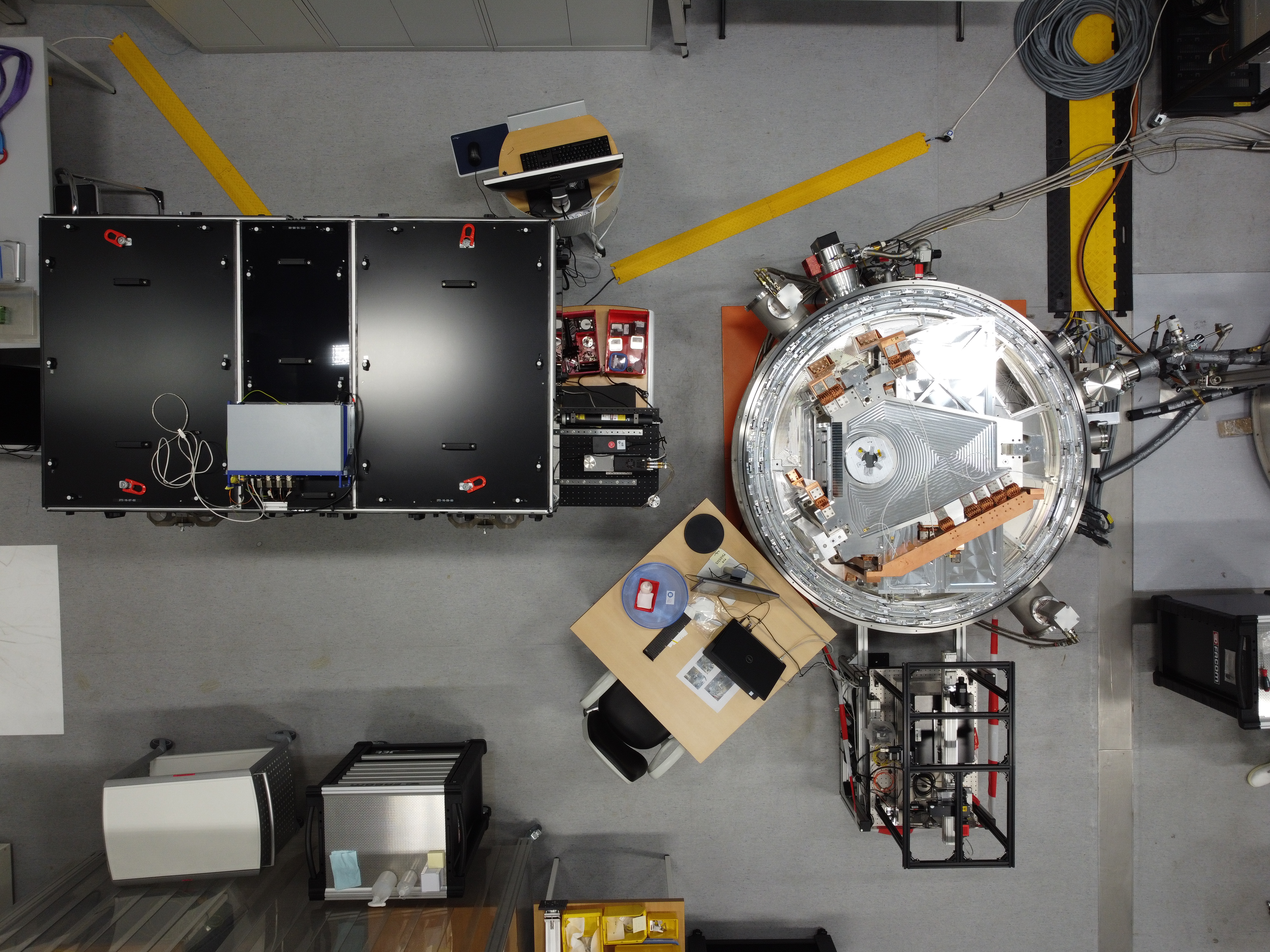}
    \vspace{1mm}
    \caption{An aerial shot of \gls*{mitesi} (left) in front of the \gls*{scao} test cryostat at in the cleanroom at \gls*{mpia} where subsystem testing of the \gls*{scao} system will be carried out.}
    \label{fig:MITESI-drone}
\end{figure}

The \gls*{mitesi} testbed has now been assembled and delivered to the cleanroom at \gls*{mpia} where it will be used to close the loop and help perform subsystem testing for the \gls*{metis} \gls*{scao} system. Figure~\ref{fig:MITESI-drone} shows \gls*{mitesi} on the left with its black cover panels fitted, in front of the \gls*{scao} test cryostat at \gls*{mpia}. Initial work to verify the testbeds performance has shown promising results, with the exception of the \gls*{psf} quality on the \gls*{psf} monitoring system camera. 

We will continue to investigate the beam quality on the \gls*{psf} monitoring system camera. We are considering further validation experiments with the test bench such as verifying the lateral shifting of the pupil images using the \gls*{fp}1 and \gls*{fp}2 mirrors, potentially with the lens attached to the \gls*{psf} monitoring system camera to reimage the pupil onto the detector. We would also validate the pointing and dynamic motion of the tip-tilt mirror located at \gls*{pp}3 either by use of an autocollimator or a camera placed at the exit focal plane \gls*{fp}3, which can trace the motion of the point source caused by \gls*{pp}3. 

In addition to validating the hardware, we are in the process of establishing control software using Catkit2 \cite{emiel_h_por_2026_20843325} to operate the testbed. 

Finally, we must still align \gls*{mitesi} to the \gls*{scao} module. The optical interface between the \gls*{scao} module and \gls*{mitesi} is a focal plane outside of the the \gls*{scao} module test cryostat. Alignment is foreseen by bringing the exit focal plane of \gls*{mitesi}, \gls*{fp}3, to the \gls*{scao} module's interface focal plane. To achieve this, for coarse adjustment the \gls*{mitesi} support frame can be moved around on built in wheels. For fine adjustment, the \gls*{mitesi} optical bench sits atop four adjustment units that individually move in X,Y and Z, allowing the entire optical bench to be finely adjusted in X,Y,Z and rotation X,Y,Z.

Using \gls*{mitesi} to test the \gls*{scao} system is scheduled to begin later this year.

\acknowledgments 

Thank you to M.Feldt for providing the simulated \gls*{elt} \gls*{psf} used in section~\ref{sec:PSF-comparison}. 
Author M. C. Cárdenas Vázquez acknowledges financial support from the foundation “MERAC” for a two-year project under the framework of its Postdoctoral Studies.

\bibliography{report} 
\bibliographystyle{spiebib} 

\end{document}